\documentclass{elsart}
\usepackage{amssymb}
\large
\begin{document}
\begin{frontmatter}
\title{Explicit integration of one problem of motion of the generalized
Kowalevski top}

\author{M.P. Kharlamov\corauthref{cor1}},
\ead{mharlamov@vags.ru}
\author{A.Y. Savushkin}

\corauth[cor1]{Corresponding author.}
\address{Chair of Mathematical Simulation, Volgograd Academy of
Public Administration, Gagarin Street 8, Volgograd, 400131, Russia}

\journal{Mechanics Research Communications} \volume{32}
\pubyear{2005} \setcounter{firstpage}{547}
\setcounter{lastpage}{552}

\date{}
\begin{abstract}
In the problem of motion of the Kowalevski top in a double force
field the \mbox{4-}dimen\-sional invariant submanifold of the phase
space was pointed out by M.P.~Kharlamov (Mekh. Tverd. Tela, {\bf
32}, 2002). We show that the equations of motion on this manifold
can be separated by the appropriate change of variables, the new
variables $s_1, s_2$ being elliptic functions of time. The natural
phase variables (components of the angular velocity and the
direction vectors of the forces with respect to the movable basis)
are expressed via $s_1, s_2$ explicitly in elementary algebraic
functions.
\end{abstract}

\end{frontmatter}

\section{Introduction}

The famous solution of S.~Kowalevski \cite{1} for the motion of a
heavy rigid body about a fixed point was generalized for the case
of double constant force field in \cite{2,3}. This Hamiltonian
system essentially has three degrees of freedom, and hardly can
receive a clear geometrical or mechanical description of all types
of motions. Invariant subsystems, which can be interpreted as
systems with two degrees of freedom, were found in \cite{2,4}.
Phase topology of the case \cite{2} was studied in \cite{5}. The
present paper deals with the case \cite{4}. We show that by proper
choice of local coordinates it is possible to obtain separated
differential equations of motion, and express all phase variables
explicitly in terms of two new variables, the latter being
elliptic functions of time.

\section{Equations of motion and known first integrals}

Consider a heavy magnetized rigid body with a fixed point placed
in gravitational and magnetic constant force fields. Let
$\vec\alpha ,\vec\beta$ be the direction vectors of the force
fields and $\vec e _1, \vec e _2$ be the radius vector of mass
center and the vector of magnetic moment of the body. The scalar
characteristics (for example, the product of weight and distance
from the mass center to the fixed point) may be included in the
length of either vector of the associated pair. We prefer to
consider $\vec e _1, \vec e _2$ to be unit vectors according to
the model accepted in \cite{3}. The Euler--Poinsot equations of
motion have the form
$$
\displaystyle{I\frac{d\vec\omega}{dt} =I\vec\omega \times
\vec\omega + \vec e _1 \times \vec \alpha + \vec e _2 \times \vec
\beta, \quad \frac {d\vec \alpha}{dt} = \vec \alpha \times \vec
\omega ,\quad \frac {d\vec \beta}{dt} = \vec \beta \times \vec
\omega,} \eqno(1)
$$
where $\vec\omega$ is the angular velocity, $I$ is the inertia
tensor. All vector or tensor objects are referred to the basis
attached to the body.

Suppose that the body is dynamically symmetric and denote by
$\pi_e$ the equatorial plane of inertia ellipsoid. Choosing
measure units one can always make the moment of inertia with
respect to symmetry axis equal 1. Let
$$
\begin{array}{c}
I =\texttt{diag} \{ 2,2,1 \} , \\
\vec e _1, \vec e _2 \in \pi_e,\quad \vec e _1 \cdot \vec e _2=0.
\end{array} \eqno (2)
$$
Then $\vec e _1 ,\vec e _2$ may be taken as the first vectors of
the movable basis. It is known that under conditions (2) the
system (1) is completely integrable due to the existence of the
first integrals \cite{2}
$$
\begin{array}{c}
H = \omega _1^2 + \omega _2^2 + \frac{1} {2}\omega _3^2 - (\alpha _1 + \beta _2 ), \\
K = (\omega _1^2 - \omega _2^2 + \alpha _1 - \beta _2 )^2 +
(2\omega _1 \omega _2 + \alpha _2 + \beta _1 )^2,
\end{array}\eqno(3)
$$
and a new integral $G$ pointed out in \cite{3}, which in case
$\vec \beta = 0$ turns into the square of the vertical component
of the angular momentum.

Below we exclude the case $ | \vec\alpha | = | \vec\beta |, \;
\vec\alpha \cdot \vec\beta = 0$, for which there exists a cyclic
coordinate \cite{3a}, and the problem reduces to the system with
two degrees of freedom.

First, we show that without loss of generality one can always take
$\vec\alpha  \bot \vec\beta$.

The conditions (2) hold if we replace $\vec e _1 ,\vec e _2
,\vec\alpha ,\vec\beta$ with $\vec e _1 (\theta),\vec e _2
(\theta),\vec\alpha (\theta),\vec\beta(\theta)$, where
$$
\begin{array}{c}
(\vec e _1 (\theta),\vec e _2 (\theta)) = (\vec e _1 ,\vec e
_2)\Theta, \quad (\vec \alpha (\theta),\vec \beta (\theta)) =
(\vec \alpha,\vec \beta)\Theta,\\
\Theta = \left( \begin{matrix} {{\cos \theta} & { - \sin \theta}
\cr {\sin \theta} & {\cos \theta}\cr }
\end{matrix}
\right), \quad \theta = \texttt{const}.
\end{array}
\eqno(4)
$$
Therefore, $\vec e _1 (\theta),\vec e _2 (\theta)$ may be taken as
the first vectors of a new movable basis. At the same time,
substitution (4) preserves the rotating moment $\vec e _1 \times
\vec \alpha + \vec e _2 \times \vec \beta$ in Euler equations, and
new vectors $\vec \alpha (\theta),\vec \beta (\theta)$ satisfy
Poisson equations.

For the general case
$$(| \vec\alpha |^2 - |\vec\beta |^2 )^2
+ (\vec\alpha \cdot\vec\beta )^2 \ne 0
$$
take ${\tan 2\theta = 2(\vec\alpha,\vec\beta)/(|\vec\alpha |^2 - |
\vec\beta |^2 )}$ if $|\vec\alpha | \ne | \vec\beta  |$, and $\cos
2\theta = 0$ if $| \vec\alpha  | = | \vec\beta |$. Then
$\vec\alpha(\theta) \bot \vec\beta (\theta)$. Thus, below we
consider the natural restrictions (also called the geometrical
integrals) in the form
$$
| \vec \alpha |^2 = a^2 ,\; | \vec \beta |^2 = b^2 ,\;\vec \alpha
\cdot \vec \beta = 0. \eqno(5)
$$

The first integral \cite{3} can then be written in a simple way
$$
G = \frac{1}{4}(\omega_\alpha^2 + \omega_\beta^2 ) +
\frac{1}{2}\omega_3\omega_\gamma - b^2 \alpha_1 - a^2 \beta_2,
\eqno(6)
$$
where
$$
\begin{array}{l}
\omega_\alpha = 2\omega_1 \alpha_1 + 2\omega_2 \alpha_2 +
\omega_3 \alpha_3 , \\
\omega_\beta = 2\omega_1 \beta_1 + 2\omega_2 \beta_2 + \omega_3 \beta_3 , \\
\omega_\gamma = 2\omega_1 (\alpha_2 \beta_3 - \alpha_3 \beta_2 ) +
2\omega_2 (\alpha_3 \beta_1 - \alpha_1 \beta_3 ) + \omega_3
(\alpha_1 \beta_2 - \alpha_2 \beta_1 ).
\end{array}
$$

The conditions (5) make the phase space diffeomorphic to $M^6 =
{\bf R}^3 \times SO(3)$. In general, the integral manifold
$J_{h,k,g} = \{ H = h, K = k, G = g \} \subset M^6$ consists of
3-dimensional tori bearing quasiperiodic trajectories dense on
each torus for almost all values of the integral constants.
Therefore, a 4-dimensional invariant submanifold, on which the
induced system has a structure of the integrable system with two
degrees of freedom, must reside in the set of critical points of
the global integral mapping $J = H \times K \times G: M^6 \to {\bf
R}^3 $ . One submanifold of this type was found in \cite{2}:
$$
M^4 = \{ K = 0\} \subset M^6.
$$
The topological structure of the induced system on $M^4 $ was
studied in \cite{5}.

Below we deal with the case \cite{4}, which generalizes the
so-called marvellous motions of the 2nd and 3rd classes of
Appelrot \cite{6}.
\section{New equations for the integral manifolds}
Changing, if necessary, the order of vectors in the movable basis
we can assume that $a > b$. Denote
$$
p^2 = a^2 + b^2 ,r^2 = a^2 - b^2
$$
and consider the set $N^4 \subset M^6$ of critical points of the
function
$$ F = (2G - p^2 H)^2 - r^4 K$$
belonging to the level $F = 0$.

In order to obtain simple formulae and to establish the
correspondence with \cite{4}, introduce new phase variables ($i^2
= - 1$)
$$
\begin{array}{c}
w_1 = \omega_1 + i\omega_2,\quad w_2 = \bar w_1,\quad w_3=\omega_3,\\
x_1 = (\alpha_1 - \beta_2) + i(\alpha_2 + \beta_1),\qquad x_2 = \bar x_1,\\
y_1 = (\alpha_1 + \beta_2) + i(\alpha_2 - \beta_1),\qquad y_2 = \bar y_1,\\
z_1 = \alpha_3 + i\beta_3, \qquad z_2 = \bar z_1.
\end{array}
\eqno(7)
$$
The equations (1) can be then written as follows
$$
\begin{array}{c}
{2w'_1 =-(w_1 w_3 + z_1),\quad 2w'_2=w_2 w_3 + z_2,\quad 2w'_3=y_2
- y_1,}\\
{x'_1 = - x_1 w_3 + z_1 w_1,\qquad x'_2 = x_2 w_3 - z_2 w_2,} \\
{y'_1 = - y_1 w_3 + z_2 w_1,\qquad y'_2 = y_2 w_3 - z_1 w_2,} \\
{2z'_1 = x_1 w_2 - y_2 w_1,\qquad 2z'_2 = - x_2 w_1 + y_1 w_2.}\\
\end{array}
\eqno(8)
$$
Here stroke stands for $d/d(it)$.

The conditions (5) take the form
$$
\begin{array}{c}
z_1^2 + x_1 y_2 = r^2,\quad z_2^2 + x_2 y_1 = r^2, \\
x_1 x_2 + y_1 y_2 + 2z_1 z_2 = 2p^2,
\end{array} \eqno(9)
$$
and $F = 0,\nabla _6 F = 0$ give
$$
F_1 = 0, \quad F_2 = 0, \eqno(10)
$$
where
$$
\begin{array}{l}
\displaystyle {F_1 = \sqrt {x_1 x_2 } w_3 - \frac{1} {{\sqrt {x_1
x_2} }}(x_2 z_1
w_1 + x_1 z_2 w_2 ),} \\
\displaystyle {F_2 = \frac{i} {2}[\frac{{x_2 }} {{x_1 }}(w_1^2 +
x_1) - \frac{{x_1}} {{x_2}}(w_2^2 + x_2)]}.
\end{array}
$$
The equations (10) correspond to the system of invariant relations
found in \cite{4}. This fact, in particular, reveals the
topological nature of strictly analytical results \cite{4}.
Moreover, it straightforwardly proves that $N^4$ is a subset of
the phase space invariant under the flow of the dynamical system
(1).

It is easy to see that almost everywhere on $N^4$ the system of
equations (10) has rank 2, so $N^4$ has a natural structure of
4-dimensional manifold except, maybe, for a thin subset defined by
$x_1 x_2 = 0$. Fix the constants $h,k,g$ of the integrals (3), (6)
and introduce new constants
$$
m = \frac{1}{{r^4}}(2g - p^2 h),\quad \ell = \sqrt {2p^2 m^2 + 2hm
+ 1}
$$
(the sign of $\ell$ is arbitrary). Then from the first integrals,
conditions (9) and equations (10) we obtain on $N^4$
$$
\begin{array}{c}
\displaystyle {w_1^2 = \frac{{x_1 }} {{x_2 }}r^2 m - x_1, \quad
w_2^2 =
\frac{{x_2 }} {{x_1 }}r^2 m - x_2,} \\
\displaystyle {w_3 = \frac{{z_1 w_1 }} {{x_1 }} + \frac{{z_2 w_2
}} {{x_2 }}},
\\
m(x^2 + z^2 ) - \ell x + \sqrt {r^4 m^2 - r^2 m(x_1 + x_2 ) + x^2
} = 0,
\end{array} \eqno(11)
$$
where
$$
x^2 = x_1 x_2 \geqslant 0,\quad z^2 = z_1 z_2 \geqslant 0.
\eqno(12)
$$
The square root in (11) equals $w_1 w_2 $, and therefore is
non-negative.

The equations (11) of the integral manifold $ J_{h,k,g}\subset
N^4$ show that in general case for given $m,\ell$ this manifold is
two-dimensional.
\section{Separation of variables}
We now introduce new variables in $(x,z)$-plane
$$
s_1 = \frac{{x^2 + z^2 + r^2 }} {{2x}},\quad s_2 = \frac{{x^2 +
z^2 - r^2 }} {{2x}}. \eqno(13)
$$
Calculating time derivatives from (8) we obtain
$$
\begin{array}{l}
\displaystyle {s'_1 - s'_2 = \frac{{r^2 }} {{2x^2 }}[z_2 \sqrt
{\frac{{x_1 }}
{{x_2 }}} w_2 - z_1 \sqrt {\frac{{x_2 }} {{x_1 }}} w_1 ],}\\
\displaystyle {s'_1 + s'_2 = \frac{{r^2 }} {{2x^2 }}[z_1 \sqrt
{\frac{{x_1 }} {{x_2 }}} w_2 - z_2 \sqrt {\frac{{x_2 }} {{x_1 }}}
w_1 ]}.
\end{array} \eqno(14)
$$
Let
$$
\begin{array}{l}
\Psi (s_1 ,s_2 ) = 4ms_1 s_2 - 2\ell (s_1 + s_2 ) +
\displaystyle{\frac{1}{m}}(\ell ^2 - 1), \\
\Phi (s) = 4ms^2 - 4\ell s +\displaystyle{\frac{1}{m}}(\ell ^2 -
1).
\end{array}
$$

Then, having the obvious identity
$$
\Psi ^2 (s_1 ,s_2 ) - \Phi (s_1 )\Phi (s_2 ) = 4(s_1 - s_2 )^2,
$$
we find from (9), (11), (12)
$$
\begin{array}{l}
\displaystyle{x_1 = - \frac{{r^2 }} {{2(s_1 - s_2)^2}}[\Psi (s_1
,s_2 ) + \sqrt
{\Phi (s_1 )\Phi (s_2 )} ],}\\
\displaystyle{x_2 = - \frac{{r^2 }} {{2(s_1 -
s_2)^2}}[\Psi (s_1 ,s_2 ) - \sqrt {\Phi (s_1 )\Phi (s_2 )} ],} \\
\displaystyle{y_1 = 2\frac{{(2s_1 s_2 - p^2 ) - 2\sqrt {(s_1^2 -
a^2 )(s_2^2 - b^2 )} }} {{\Psi (s_1 ,s_2 ) - \sqrt {\Phi (s_1
)\Phi (s_2 )}
}},}\\
\displaystyle{y_2 = 2\frac{{(2s_1 s_2 - p^2 ) + 2\sqrt {(s_1^2 -
a^2 )(s_2^2 - b^2 )} }} {{\Psi (s_1 ,s_2 ) + \sqrt {\Phi (s_1
)\Phi
(s_2 )} }},} \\
\displaystyle{z_1 = \frac{r} {{s_1 - s_2 }}(\sqrt {s_1^2 - a^2 } +
\sqrt {s_2^2
- b^2 } ),}\\
\displaystyle{z_2 = \frac{r} {{s_1 - s_2 }}(\sqrt {s_1^2 - a^2 }
- \sqrt {s_2^2 - b^2 } ),} \\
\displaystyle{w_1 = r\frac{{\sqrt {\Phi (s_2 )} - \sqrt {\Phi (s_1
)} }} {{\Psi
(s_1 ,s_2 ) - \sqrt {\Phi (s_1 )\Phi (s_2 )} }},}\\
\displaystyle{w_2 = r\frac{{\sqrt {\Phi (s_2 )} + \sqrt {\Phi (s_1
)} }} {{\Psi (s_1
,s_2 ) + \sqrt {\Phi (s_1 )\Phi (s_2 )} }},} \\
\displaystyle{w_3 = \frac{1} {{s_1 - s_2 }}[\sqrt {(s_2^2 - b^2
)\Phi (s_1 )} - \sqrt {(s_1^2 - a^2 )\Phi (s_2 )} ].}
\end{array} \eqno(15)
$$
Substitution of the latter expressions for $x_j ,z_j ,w_j (j =
1,2)$ into (14) allows to obtain the differential equations for
$s_1 ,s_2$ in the real form
$$
\displaystyle{\frac{ds_1}{dt} =\frac{1}{2} \sqrt {(a^2 - s_1^2
)\Phi (s_1 )}, \; \frac{ds_2}{dt} =\displaystyle{\frac{1}{2}}
\sqrt {(b^2 - s_2^2 )\Phi (s_2 )}}. \eqno(16)
$$
Thus, $s_1 ,s_2 $ are easily found as elliptic functions of time.
The explicit formulae for the phase variables $\omega_j ,\alpha_j,
\beta_j (j = 1,2,3)$ immediately follow from (7), (15).

\section{The types of solutions}
The conditions (9) define the global area for the variables (13):
$$
|s_1| \geqslant a,\quad |s_2| \leqslant b.
$$
So, the domain of oscillations (16) for the fixed values of
$m,\ell $ is obtained from the inequalities
$$
\Phi (s_1 ) \leqslant 0,\quad \Phi (s_2 ) \geqslant 0.
$$
Bifurcations of solutions (16) with respect to the parameters
$m,\ell$ take place when $\Phi (\pm a) = 0$, or $\Phi ( \pm  b) =
0$. It leads to a set of lines in $(m,\ell )$-plane
$$
\ell = -2am \pm 1,\;\ell = 2am \pm 1,\;\ell = - 2bm \pm 1,\;\ell =
2bm \pm 1.
$$

Analyzing the evolution of roots of the polynomial $\Phi (s)$, we
obtain all different types of motions.

The critical motions appear in the cases when one of the variables
$s_1 ,s_2 $ remains constant, coinciding with the double root of
polynomial product in the right part of the corresponding equation
(16). It obviously leads to the motion of the body of pendulum
type: either $\omega _1 = \omega _3 \equiv 0$, or $\omega _2 =
\omega _3 \equiv 0$.

An interesting aspect of the case considered is that both movable
and immovable hodographs of the angular velocity are explicitly
found simultaneously without any further integration. Actually,
because for almost all constants $m,\ell $ from the domain of
existing of real solutions, hodographs fill some two-dimensional
surfaces densely, we use the expressions (15) to obtain the
parametric equations of those surfaces, in which $s_1 ,s_2 $ are
independent parameters. Expressions for $\vec\alpha, \vec\beta$
(and, therefore, for $\vec\alpha \times \vec\beta$) via $s_1, s_2$
give the parametric equations for the orientation matrix.
Contemporary methods of computer graphics provide the possibility
to construct a detailed and clear picture of motion as rolling
without slipping of one surface through the other: at any moment
these surfaces have a common point with zero absolute velocity and
one common tangent vector, expressing the fact that absolute and
relative derivatives of the angular velocity coincide.


\begin{thebibliography}{00}
\bibitem {1}
S. Kowalevski, Acta Math. {\bf 12}, 177-232 (1889).
\bibitem {2}
O.I. Bogoyavlensky, Commun. Math. Phys. {\bf 95}, 307-315 (1984).
\bibitem {3}
A.I. Bobenko, A.G. Reyman, M.A. Semenov-Tian-Shansky, Commun.
Math. Phys. {\bf 122}, 321-354 (1989).
\bibitem {4}
M.P. Kharlamov, Mekh. Tverd. Tela, {\bf 32}, 32-38 (2002).
\bibitem {5}
D.B. Zotev, Regular and Chaotic Dynamics, {\bf 5} (4), 437-458
(2000).
\bibitem {3a}
H.M. Yehia, Mech. Res. Commun. {\bf 13} (3), 169-172 (1986).
\bibitem {6}
G.G. Appelrot, Non-completely symmetric heavy gyroscopes, in
"Motion of the rigid body about a fixed point", 1940,
Moscow-Leningrad, 61-156.
\end{thebibliography}
\end{document}